\documentclass{aa}  
\usepackage{graphicx,array,amsmath,amssymb}
\usepackage[authoryear]{natbib}
\bibpunct[]{(}{)}{;}{a}{}{,}

\begin{document}

\def\gtapp
{\mathrel{\hbox{\raise0.3ex\hbox{$>$}\kern-0.8em\lower0.8ex\hbox{$\sim$}}}}
\def\ltapp
{\mathrel{\hbox{\raise0.3ex\hbox{$<$}\kern-0.75em\lower0.8ex\hbox{$\sim$}}}}

\title{Are the hosts of Gamma-Ray Bursts sub-luminous and blue
       galaxies\,?\,\thanks{Based on observations with the Very Large
       Telescope, obtained at the European Southern Observatory in
       Chile under Proposal 67.B-0611(A)}$^,\,$\thanks{Based on observations with the Gemini-North
       Telescope, obtained at Mauna Kea (Hawaii) under Proposal GN-2001A-Q-58}}
  
\author{E.~Le~Floc'h \inst{1}  
\and P.-A.~Duc \inst{1,2} 
\and I.F.~Mirabel \inst{1,3}
\and D.B.~Sanders \inst{4,5}  
\and G.~Bosch \inst{6}
\and R.J.~Diaz \inst{7}
\and C.J.~Donzelli \inst{8}
\and I.~Rodrigues \inst{1}
\and T.J.-L.~Courvoisier \inst{9,10}
\and J.~Greiner \inst{5}
\and S.~Mereghetti \inst{11}
\and J.~Melnick \inst{12}
\and J.~Maza \inst{13}
\and D.~Minniti \inst{14}
}
  
\institute{CEA/DSM/DAPNIA Service d'Astrophysique, F-91191 Gif-sur-Yvette,  
France  
\and   
CNRS URA 2052, France
\and
Instituto de Astronom\'\i a y F\'\i sica del Espacio, cc 67, suc 28.   
1428 Buenos Aires, Argentina
\and
Institute for Astronomy, University of Hawaii, 2680 Woodlawn Drive, 
Honolulu, HI 96822, United States
\and  
Max-Planck-Institut f\"ur Extraterrestrische Physik, D-85740, Garching, Germany
\and
Facultad de Cs. Astronomicas y Geof\'{\i}sica, Paseo del Bosque s/n, La Plata, Argentina
\and
Observatorio Astron\'omico de C\'ordoba \& SeCyT, UNC, Laprida 854, Cordoba
[5000], Argentina
\and
IATE, Observatorio Astron\'omico \& CONICET, Laprida 854, Cordoba [5000], Argentina 
\and
INTEGRAL Science Data Center, Ch. d'Ecogia 16, CH-1290 Versoix, Switzerland
\and
Geneva Observatory, Ch. des Maillettes 11, 1290 Sauverny, Switzerland
\and
Istituto di Astrofisica Spaziale e Fisica Cosmica, Sezione di Milano ``G. Occhialini'', 
via Bassini 15, I-20133 Milan, Italy 
\and
European Southern Observatory, Alonso de Cordova 3107, Santiago, Chile
\and 
Departamento de Astronom\'{\i}a, Universidad de Chile, Casilla 36-D, Santiago, Chile
\and
Department of Astronomy, Pontifica Universidad Cat\'olica, Av. Vicu\~na Mackenna 4860, Santiago, Chile
} 
  
\titlerunning{Are the GRB hosts sub-luminous and blue galaxies\,?}
\authorrunning{Le Floc'h et al.}
  
\offprints{E. Le Floc'h (elefloch@cea.fr)}  
  
\date{Received December 6, 2002 / Accepted December 23, 2002}  
  
\abstract{ We present $K$--band imaging observations of ten Gamma-Ray
Burst (GRB) host galaxies for which an optical and/or radio afterglow
associated with the GRB event was clearly identified.  Data were
obtained with the Very Large Telescope and New Technology Telescope at
ESO (Chile), and with the Gemini-North telescope at Mauna Kea
(Hawaii).  Adding to our sample nine other GRB hosts with $K$--band
photometry and determined redshifts published in the literature, we
compare their observed and absolute $K$ magnitudes as well as their
$R-K$ colours with those of other distant sources detected in various
optical, near-infrared, mid-infrared and submillimeter deep surveys.
We find that the GRB host galaxies, most of them lying at 0.5\,$\ltapp
z \ltapp$\,1.5, exhibit very blue colours, comparable to those of the
faint blue star-forming sources at high redshift.  They are
sub-luminous in the $K$--band, suggesting a low stellar mass
content. We do not find any GRB hosts harbouring $R$-- and $K$--band
properties similar to those characterizing the luminous
infrared/submillimeter sources and the extremely red starbursts.
Should GRBs be regarded as an unbiased probe of star-forming activity,
this lack of luminous and/or reddened objects among the GRB host
sample might reveal that the detection of GRB optical afterglows is
likely biased toward unobscured galaxies. It would moreover support
the idea that a large fraction of the optically-dark GRBs occur within
dust-enshrouded regions of star formation.  On the other hand, our
result might also simply reflect intrinsic properties of GRB host
galaxies experiencing a first episode of very massive star formation
and characterized by a rather weak underlying stellar population.
Finally, we compute the absolute $B$ magnitudes for the whole sample
of GRB host galaxies with known redshifts and detected at optical
wavelengths. We find that the latter appear statistically even less
luminous than the faint blue sources which mostly contributed
to the $B$--band light emitted at high redshift. This indicates that
the formation of GRBs could be favoured in particular systems with
very low luminosities and, therefore, low metallicities. Such an
intrinsic bias toward metal-poor environments would be actually
consistent with what can be expected from the currently-favoured
scenario of the ``collapsar''.  The forthcoming launch of the SWIFT
mission at the end of 2003 will provide a dramatic increase of the
number of GRB-selected sources.  A detailed study of the chemical
composition of the gas within this sample of galaxies will thus allow
us to further analyse the potential effect of metallicity in the
formation of GRB events.

\keywords{
	galaxies: starburst --
	galaxies: evolution --
	cosmology: observations --
	gamma rays: bursts --
	}  
}  

\maketitle
  
\section{Introduction}  

In the past few years, a variety of high redshift star-forming sources
such as the Lyman- and Balmer-break galaxies \citep[e.g.][]{Steidel99}
or the SCUBA and ISO dusty starbursts
\citep{Barger98,Aussel99,Elbaz99} were discovered at different
wavelengths from a broad range of observing techniques.  However, each
of these methods of detection strongly suffers from its own selection
effects (e.g., dust extinction, flux limitation or colour selection at
the observed wavelength), and the connection between the various
populations of distant objects currently known is still poorly
understood \citep[e.g.,][]{Webb03}.

In the goal to trace the star-forming activity and the evolution of
high redshift galaxies with reduced biases, an alternative approach
using the cosmological Gamma-Ray Bursts (GRBs) as probes of star
formation was recently proposed
\citep[e.g.,][]{Wijers98,Mirabel00,Blain00}.  Since the discovery of
X-ray/optical/radio transient counterparts to long-duration GRBs, the
``collapsar'' model \citep{Woosley93,MacFadyen99} linking these events
to the cataclysmic destruction of massive stars has indeed received a
strong support from a growing number of evidence. These include the
presence of dust extinction in the X-ray and optical transients
\citep[e.g.,][]{Galama01}, the spectral energy distribution and
morphology of GRB hosts consistent with compact, irregular or
merger-driven starbursts
\citep[e.g.,][]{Hjorth02,Sokolov01,Djorgovski98}, and the GRB
localizations within their hosts suggesting a population of
disk-residing progenitors \citep{Bloom02a}.  Moreover, the iron
emission lines detected in their X-ray afterglows
\citep[e.g.,][]{Piro00} and the late-time brightenings observed in the
light-curves of several GRB optical transients
\citep[e.g.,][]{Galama00} have been interpreted as the signature for
the presence of an underlying supernova occuring with the GRB explosion,
and thus provided further clues for the ``collapsar'' scenario.
Because of the short-lived nature of massive stars, and because
gamma-rays do not suffer from intervening columns of gas and dust,
GRBs could thus be used to sign-post the instantaneous star formation
in the Universe independently of the effects of dust extinction.

In this perspective, it is especially crucial to establish whether
GRB-selected galaxies are really representative of the star-forming
sources in the field at similar redshifts \citep{Schaefer00}, or
whether they may form a particular category of new objects or any
sub-sample of an already-known population of galaxies.  Physical
properties of the circum-burst environment could play a crucial role
in the formation of these cataclysmic events.  For example, GRBs could
be favoured in low-metallicity regions of star formation
\citep{MacFadyen99}, and we would expect to detect these phenomena
preferentially in dwarf and sub-luminous galaxies.  On the other hand,
their association with the destruction of massive stars may imply that
GRBs are mostly observed in environments experiencing powerful
episodes of star formation such as the luminous starburst galaxies.

Here we report on our GRB host imaging program carried out in the
near-infrared (NIR) at European Southern Observatory (ESO) and Gemini
Observatory.  This program complements the $K$--band data of GRB host
galaxies already obtained in the northern hemisphere
\citep[e.g.,][]{Chary02}.  Observations performed in the NIR present
multiple interests.  They first allow to probe the light emitted by
the old star populations, which provides a good indication of the
stellar mass of galaxies.  For high redshift sources, the relative
importance of the NIR versus optical emission can moreover be used to
roughly estimate the level of dust obscuration.  Finally, the
selection of distant objects in the NIR is nearly insensitive to the
spectral energy distribution (SED) of galaxies up to very high
redshift (z~$\sim$~2) because of an invariant k-correction along the
Hubble sequence.

We describe our observations and data reduction in
Sect.\,~\ref{sec:red}, and present our results in
Sect.\,\ref{sec:results}. We discuss these new data in
Sect.\,\ref{sec:discuss} and finally conclude in
Sect.\,\ref{sec:conclusion}. Additional information
on the GRB host optical properties are mentioned
in Appendix\,A.

\begin{table*}[htbp]
\caption{Summary of observations}
\begin{center}
\begin{tabular}{ccccccc}
\hline \hline
  &       &      &    & T$_{obs}$--T$_{grb}~(\dag)$  & On-Source  &  Seeing \\
  & Source~$^\P$ & GRB~$^\amalg$ &  &   (days) 	      & Time (s)    & (\arcsec) \\

\hline \\
\multicolumn{2}{l}{\it ISAAC observations  } & & & & &                    \\
& GRB~J115450.1--264035 & 990506  & &  701 & 3600 & 1.05 \\
& GRB~J182304.6--505416 & 001011  & &  173 & 3600 & 0.70 \\
&  GRB~J122519.3+200611 & 000418  & &  422 & 3600 & 1.50 \\
& GRB~J015915.5--403933 & 000210  & &  509 & 3600 & 1.20 \\
& GRB~J232937.2--235554 & 981226  & &  893 & 3600 & 0.85 \\
& GRB~J061331.1--515642 & 000131  & &  599 & 3600 & 0.75 \\
& GRB~J133807.1--802948 & 990510  & &  697 & 3600 & 1.05 \\
\\									
\multicolumn{2}{l}{\it SOFI observations  } & & & & &     \\                
& GRB~J223153.1--732429 & 990712 & & 403 & 5520 & 0.60 \\
&  GRB~J122311.4+064405 & 990308 & & 362 & 4200 & 0.75 \\
\\										
\multicolumn{2}{l}{\it Hokupa'a/QUIRC observations  } & & & &             &   \\
& GRB~J070238.0+385044 & 980329 & & 1000 & 4320 & 0.15$^\ddagger$ \\
\hline \hline

\end{tabular}\\
\end{center}
\vspace{-0.2cm}
{\bf Notes:}

\hspace{.11cm}
$\P$ : host galaxies, named after their selection criteria (GRB) and their
equatorial coordinates
given in the 

\hspace{.64cm} standard equinox of J2000.0.

\hspace{.11cm}
$\amalg$ : official designation of the Gamma-Ray Burst which led to the selection of the corresponding
host galaxy.

\hspace{.2cm}
$\dag$ : number of days between the GRB event and the date of our observations of the host galaxy.

\hspace{.2cm}
$\ddagger$ : resulting from the Hokupa'a Adaptive Optics correction.

\label{tab:obs}
\end{table*}

\section{Observations, reduction and analysis}  
\label{sec:red}  

Observations were performed using the ESO facilities in Chile and the
Gemini-North telescope in Hawaii.  Ten GRB host galaxies, most of them
located in the southern hemisphere and selected for having had an
optical and/or radio bright afterglow, were imaged at near-infrared
wavelengths.  Our sample of sources is listed in Table~\ref{tab:obs}
together with a log of the observations.

\subsection{Near-infrared observations}
\label{sec:nir_obs}

The NIR data were obtained with the Infrared Spectrometer And Array
Camera (ISAAC) on the Very Large Telescope (VLT) at Paranal, the Son
OF ISAAC (SOFI) installed on the New Technology Telescope (NTT) at La
Silla, and with the Adaptive Optics Hokupa'a\,/\,QUIRC instrument on
Gemini-North at Mauna Kea.  Observations were carried out between
March 2000 and September 2001 under photometric conditions. A $K_s$
filter (2.0--2.3\,$\mu$m) was used for the ISAAC and SOFI data, while
the observations on Gemini were performed with a $K'$ filter
(1.9--2.3\,$\mu$m).  The focal lens configurations resulted in a
respective pixel size of 0{\farcs}148, 0{\farcs}297 and 0{\farcs}02
for the ISAAC, SOFI and Hokupa'a images. The seeing remained rather
stable during the observations of one given source, though it varied
between 0{\farcs}6 and 1{\farcs}5 from one night to another.
Individual frames were obtained as a co-addition of 12~single
exposures of 10\,seconds each.  The series of acquisition for each
object were then carried out in a jitter mode, with a dither of the
frames following either a random pattern characterized by typical
offsets $\sim$ 30{\arcsec} on the sky for the ISAAC and SOFI images,
or a regular grid with shifts of 5{\arcsec} for the Hokupa'a data.
For the ISAAC observations, we reached a total on-source integration
time of 1\,hour per object.

Data reduction was performed following the standard techniques of NIR
image processing.  To estimate the thermal background contribution of
each frame, a ``sky'' map was generated using a median-average of the
9 jittered images directly preceding and following a given
acquisition.  The corresponding ``sky'' was then scaled to the mode of
the object frame and subsequently subtracted. This method allowed us
to remove in the meantime the contributions of the bias and dark
current. Finally, the differential pixel-to-pixel response of the
arrays was corrected using flat-field images taken as part of the
instrument calibration plans. For the ISAAC and Hokupa'a data, these
flat-fields were obtained by observing a blank-field of the sky during
twilights, while a white screen within the dome of the NTT was used
for the SOFI observations. For the latter, we noticed that the low
spatial frequencies of the detector sensitivity were not properly
taken into account with the dome images. They were therefore corrected
using a low-order polynomial 2D-fit of the array response, a method
often refered as the ``\,illumination correction technique\,''.
Photometric calibrations were performed using the NICMOS standard
stars from \citet{Persson98}.

\begin{figure*}[htbp]
\centering
\includegraphics[width=16cm]{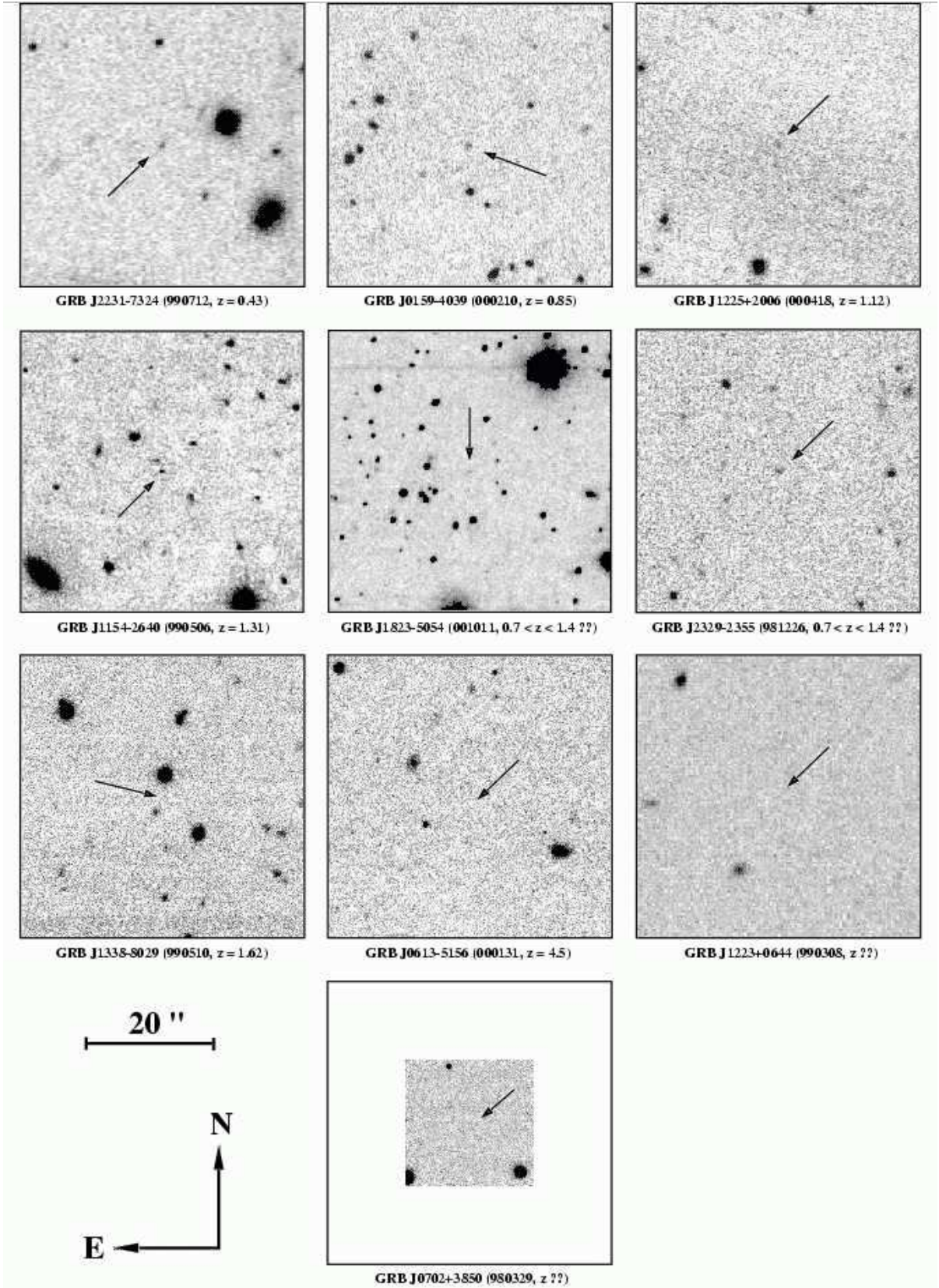}
\caption{Near-infrared images of the Gamma-Ray Burst host galaxies
listed in Table\,\ref{tab:obs}. Observations were performed with a
$K_s$ filter, except in the case of the GRB\,980329 host for which a
$K'$ filter was used.  Each frame has a field of view of
45\,{\arcsec}$\times$\,45\,{\arcsec} and an orientation with the North
to the top and the East to the left.  From the top to the bottom and
the left to the right, they were tentatively ordered with increasing
distance of the host from Earth.  The GRB\,981226 and GRB\,001011 host
galaxies have been assigned a plausible redshift range as described in
Sect.\,\ref{sec:mag_obs}, while references for the other redshifts are
given in Table\,\ref{tab:colours}.  In the last four images, the
targets were not detected. }
\label{fig:images}
\end{figure*}
  
\begin{table*}[htbp]
\caption{Properties of GRB host galaxies}
\begin{center}
\begin{tabular}{cccccrlccc}
\hline \hline
        &    &  \multicolumn{2}{c}{Redshift}   & & \multicolumn{5}{c}{Photometry} \\ 
\cline{3-4} 
\cline{6-10} \\
Source & GRB  & z & Ref. & & \multicolumn{1}{c}{$K$ mag} &  \multicolumn{1}{l}{References} & & 
\multicolumn{1}{c}{$R-K$ colour~$^\ddag$} & Abs. K mag.$~^\P$ \\
\hline \\
GRB  J225559.9+405553 & 010921  &  0.45	& 1  &  & 19.05 $\pm$ 0.1~       & 1         & & 2.40 $\pm$ 0.25 &  -22.50 \\
GRB  J145212.5+430106 & 010222  &  1.48	& 2  &  & 23.5  $\pm$ 0.0$^\dag$ & 3         & & 2.20 $\pm$ 0.30 &  -21.25 \\
GRB J182304.6--505416 & 001011  &    	&    &  & 21.45 $\pm$ 0.2~       & this work & & 3.75 $\pm$ 0.45 & 	  \\
GRB  J122519.3+200611 & 000418  &  1.12 & 4  &  & 21.3  $\pm$ 0.2~       & this work & & 2.50 $\pm$ 0.40 &  -22.65 \\
GRB J015915.5--403933 & 000210  &  0.85	& 5  &  & 20.95 $\pm$ 0.2~       & this work & & 2.50 $\pm$ 0.30 &  -22.25 \\
GRB J061331.1--515642 & 000131  &  4.5  & 6  &  &     $\geq$ 22.5~       & this work & &                 & 	  \\ 
GRB  J163353.5+462721 & 991208  &  0.71 & 7  &  & 21.7  $\pm$ 0.2~       & 8         & & 2.60 $\pm$ 0.40 &  -21.00 \\
GRB J223153.1--732429 & 990712  &  0.43	& 9  &  & 20.05 $\pm$ 0.1~       & this work & & 1.80 $\pm$ 0.30 &  -21.40 \\
GRB J133807.1--802948 & 990510  &  1.62 & 9  &  &     $\geq$ 22.5~       & this work & &                 & 	  \\ 
GRB J115450.1--264035 & 990506  &  1.31	& 4  &  & 21.45 $\pm$ 0.2~       & this work & & 4.05 $\pm$ 0.35 &  -22.90 \\
GRB  J122311.4+064405 & 990308  &       &    &  &     $\geq$ 21.5~       & this work & &                 & 	  \\ 
GRB  J152530.3+444559 & 990123  &  1.60	& 10 &  &  21.9 $\pm$ 0.4~       & 8, 11     & & 2.40 $\pm$ 0.80 &  -23.10 \\
GRB J232937.2--235554 & 981226  &       &    &  &  21.1 $\pm$ 0.2~       & this work & & 3.40 $\pm$ 0.50 & 	  \\
GRB  J235906.7+083507 & 980703  &  0.97	& 12 &  &  19.6 $\pm$ 0.1~       & 8         & & 2.80 $\pm$ 0.30 &  -23.95 \\
GRB  J070238.0+385044 & 980329  &       &    &  &     $\geq$ 23.0~       & this work & &                 & 	  \\ 
GRB  J115626.4+651200 & 971214  &  3.42	& 13 &  &  22.4 $\pm$ 0.2~       & 8         & & 3.20 $\pm$ 0.40 &  -24.45 \\
GRB  J180831.6+591851 & 970828  &  0.96	& 14 &  &  21.5 $\pm$ 0.3~       & 14        & & 3.60 $\pm$ 0.60 &  -22.05 \\
GRB  J065349.4+791619 & 970508  &  0.83	& 15 &  &  22.7 $\pm$ 0.2~       & 8         & & 2.40 $\pm$ 0.40 &  -20.45 \\
GRB  J050146.7+114654 & 970228  &  0.69	& 16 &  &  22.6 $\pm$ 0.3~       & 8, 17     & & 2.00 $\pm$ 0.50 &  -20.05 \\

\hline \hline
\end{tabular}\\
\label{tab:colours}
\end{center}

\vspace{-0.2cm}

{\bf Notes:}

\hspace{.2cm} $\ddagger$ : for all sources except the GRB\,981226 and
GRB\,001011 hosts, the $R-K$ colours were estimated using the $R$
magnitudes

\hspace{.64cm} given in Table\,\ref{tab:r_mag} of
Appendix\,A. The $R$--band photometry for the host galaxy of GRB\,981226 has been derived from 

\hspace{.64cm} \citet{Saracco01a}, \citet{Frail99} and \citet{Holland00b}, while that of the GRB\,001011
host is taken from 

\hspace{.64cm} \citet{Gorosabel02}.

\hspace{.13cm}
$\P$ : defined as M$_{\rm K}+5 \log_{10}$ h$_{65}$ assuming a $\Lambda$CDM Universe with $\Omega_m = 0.3$ and $\Omega_{\lambda} = 0.7$ (h$_{65}$\,=\,H$_0$\,(km\,s$^{-1}$\,Mpc$^{-1}$)\,/\,65).

\hspace{.2cm}
$\dag$ : estimated from an extrapolation of the afterglow $K$--band light curve
\citep{Frail02}.

\vspace{0.1cm}

{\bf References:} 
(1) \citealt{Price02}~; 
(2) \citealt{Jha01}~;
(3) \citealt{Frail02}~; 
(4) \citealt{Bloom02c}~;
(5) \citealt{Piro02a}~;
(6) \citealt{Andersen00}~;
(7) \citealt{Castro-Tirado01}~:
(8) \citealt{Chary02}~; 
(9) \citealt{Vreeswijk01}~;
(10) \citealt{Kulkarni99}~;
(11) \citealt{Bloom99}~; 
(12) \citealt{Djorgovski98}~;
(13) \citealt{Kulkarni98}~;
(14) \citealt{Djorgovski01a}~; 
(15) \citealt{Bloom98a}~; 
(16) \citealt{Bloom01b}~;
(17) \citealt{Fruchter99b}.
\end{table*}

\subsection{Photometry}

Each galaxy was observed more than 150~days after the date of its
hosted GRB event (see column~3 of Table~\ref{tab:obs}).  Assuming the
least favourable case of a bright GRB optical transient ($R$~mag
$\sim$ 20 at GRB\,+\,2~days) with a break in the light-curve occuring
$\sim$ 2 days after the burst and a slow decay with time (temporal
index $\beta \sim -1.5$), we estimate that all GRB counterparts should
have been fainter than $R$~mag $\sim$ 35 at the time of the
observations.  Taking account of a power law spectrum $F_{\nu}
\varpropto \nu^{-1}$ for the modelling of the afterglow emission, we
set a lower limit $K$~mag $\approx$ 33.  In our data, the flux
contribution of any extra light from the fading afterglows should
therefore be completely negligible relative to the emission of the
host galaxies.

Our final images are presented in Fig.\,1. For each observation, the
astrometry was performed using foreground stars of the USNO catalog,
and the GRB hosts were identified within 1\,{\arcsec} of the positions
of the GRB transients.  Among the ten sources of our sample, six host
galaxies are clearly detected in our $K_s$--band data.  Using the task
{\it phot} within the IRAF
package\footnote{http://iraf.noao.edu/iraf/web/}, we measured their
total magnitude in an aperture of 5\,{\arcsec} in diameter centered on
the source, with the exception of GRB\,990506 host which lies only
$\sim$\,1.8\,{\arcsec} from another extended object. Since this host
galaxy has a very compact morphology at optical wavelengths
(FWHM\,$\sim$\,0.14\,{\arcsec}), as revealed by the high resolution
HST images \citep{Holland00d}, we assumed that we get a good
estimation of its overall emission within an aperture of
$\sim$\,1.5\,{\arcsec} in diameter, inside which $\sim$\,95\% of the
total flux would be included if the light profile is gaussian. Given
our typical uncertainty on the photometry ($\sim$\,0.2~mag) and taking
account of the prescriptions mentioned in the ISAAC Data Reduction
Guide\footnote{http://www.eso.org/instruments/isaac/drg/html/drg.html},
we found the $K-K_s\,$ colour terms to be negligible in the final
conversions to $K$ magnitudes.

The foreground Galactic extinctions in the direction of our targets
were derived from the DIRBE/IRAS dust maps of \citet{Schlegel98}
assuming the R$_V$\,=\,3.1 extinction curve of \citet{Cardelli89}.
The final dereddened magnitudes of our sources are given in
Table~\ref{tab:colours}, together with their redshifts obtained from
various papers of the literature.  To increase the size of our sample,
we also added in our analysis nine other GRB hosts with a determined
$K$--band photometry already published by other authors (see caption
of Table\,\ref{tab:colours} for references).  Including our results,
the number of GRB host galaxies detected in the NIR by October~2002
thus amounts to 15~sources\footnote{We did not consider the case of
GRB\,980613. In spite of the $K$--band detection of its complex
host-environment by \citet{Djorgovski00}, the $K$~magnitude of its
true host \citep[component H, see][]{Hjorth02} has not been determined
so far.} out of the $\sim$\,35 GRBs which were so far localized with a
sub-arcsecond error box \citep{Greiner02}.

\section{Results}
\label{sec:results}

The GRB host galaxies span a broad range of redshifts (see
Table\,\ref{tab:r_mag} of Appendix\,A), but the current sample of
these GRB-selected sources is actually too small to study the
evolution of their characteristics with different lookback times.  On
the other hand, it can be particularly interesting to consider these
objects as a whole sample of high-z sources, and compare their
properties with other field galaxies selected by different observing
techniques. In this section, we compare the observed and absolute
magnitudes of the GRB host galaxies at different wavelengths (e.g.,
$B$--, $R$--, $K$--band) as well as their $R-K$ colours, with those of
high redshift sources detected in the optical, NIR, mid-infrared and
submillimeter deep surveys.

\subsection{``$K$ -- z'' and ``$R$ -- z'' diagrams}
\label{sec:mag_obs}

\begin{figure}[htbp]
\resizebox{\hsize}{!}{\includegraphics{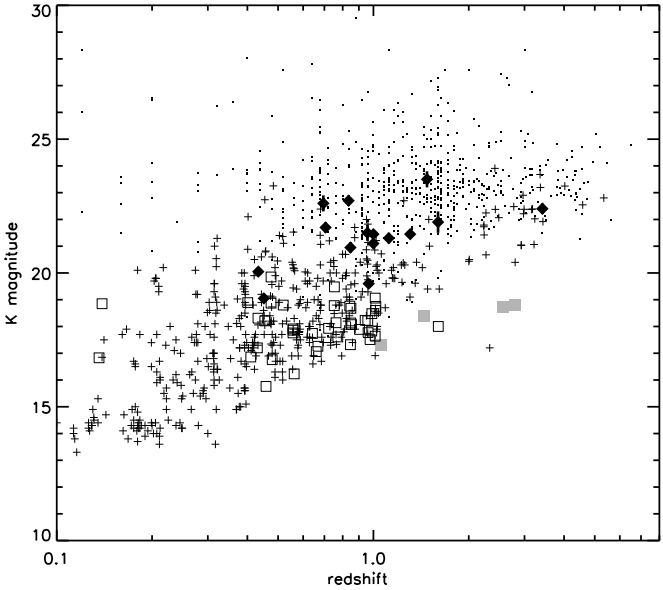}}
  \caption{Observed $K$ magnitudes of the GRB host galaxies versus
  redshift ({\it filled diamonds}) derived from
  Table\,\ref{tab:colours}.  The GRB\,001011 and GRB\,981226 hosts are
  indicated assuming an arbitrary redshift z\,=\,1.  Photometric
  uncertainties are reported with vertical solid lines.  The
  $K$~magnitudes of NIR-selected field sources with spectroscopic
  ({\it crosses}) and photometric ({\it dots}) redshifts are given for
  comparison (see Sect.\,\ref{sec:mag_obs} for references).  We also
  indicate the $K$--band photometry of the NIR counterparts to high
  redshift ISO galaxies ({\it open squares}) and SCUBA sources with
  determined spectroscopic redshifts ({\it filled squares}).  }
\label{fig:k_z}
\end{figure}  

\begin{figure}[htbp]
\resizebox{\hsize}{!}{\includegraphics{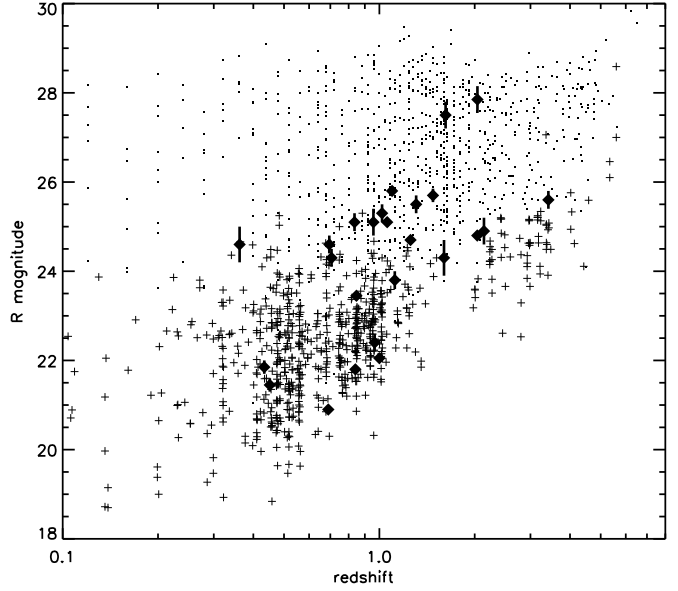}}
  \caption{Observed $R$ magnitudes of the GRB host galaxies versus
  redshift ({\it filled diamonds}) derived from
  Table\,\ref{tab:r_mag}. The uncertainties are indicated with
  vertical solid lines.  The largest ones reflect, for a given object,
  the scatter of the various magnitudes given in the literature (see
  Appendix\,A). The $R$~magnitudes of optically-selected field sources
  with spectroscopic ({\it crosses}) and photometric ({\it dots})
  redshifts are given for comparison (see Sect.\,\ref{sec:mag_obs} for
  references).}
\label{fig:r_z}
\end{figure}  

The Hubble ``$K$--z'' diagram of the GRB host galaxies is illustrated
in Fig.\,\ref{fig:k_z}.  The spectroscopic redshifts of the
GRB\,981226 and GRB\,001011 hosts have not been so far determined.
Based on their $K$ magnitude and $R-K$ colour (see
Sect.\,\ref{sec:colours}), we estimate that these objects could be
located in the 0.7\,$\ltapp z \ltapp$\,1.4 redshift range.  To allow
comparisons with other sources in the field, we overplotted the $K$
magnitudes of galaxies reported from various surveys. Nearby sources
were taken from the Hawaii $K$--band galaxy survey
\citep[$\bar{z}$\,=\,0.35,][]{Cowie94,Songaila94}, while galaxies at
intermediate redshift ($\bar{z}$\,=\,0.8) were derived from the Hawaii
Deep Surveys by \citet{Cowie96}.  Those at higher z
($\bar{z}$\,=\,1.5) were taken from the catalog of photometric
redshifts in the Hubble Deep Field \citep[HDF,][]{Fernandez_Soto99}.
We also indicated the $K$ magnitudes of the ISO sources observed in
the CFRS and HDF with flanking fields as given by \citet{Flores99},
\citet{Hogg00} and \citet{Cohen00}, as well as those of the NIR
counterparts to the SCUBA sources with confirmed spectroscopic
redshifts, obtained by \citet{Smail02} and \citet{Dey99}. The
$K$--band luminosities of these SCUBA galaxies were de-magnified from
gravitational amplification for the lensed cases.

The comparison suggests that in the NIR, the GRB host galaxies do not
particularly distinguish themselves from the field sources selected in
optical/NIR deep surveys. No particular bias of detection toward the
luminous sources is in fact apparent. There is however a significant
contrast between their $K$ magnitudes and those of the ISO and SCUBA
sources which, like the GRB hosts, are birthplaces of massive star
formation.  These differences of magnitudes and the implications on
their absolute luminosities will be more firmly established in
Sect.\,\ref{sec:mag_abs} and discussed in Sect.\,\ref{sec:discuss}.

To further address the nature of galaxies selected by GRBs relative to
other field sources at high redshift, we also present in
Fig.\,\ref{fig:r_z} \,the Hubble ``$R$--z'' diagram for the sample of
GRB hosts detected at optical wavelengths.  Their $R$ magnitudes are
given in Table\,\ref{tab:r_mag}.  They were obtained from various
papers of the literature and homogenized following the method
described in Appendix\,A. This sample is significantly larger than the
one selected in the $K$--band. In addition to the hosts which have not
been imaged in the NIR yet, there is indeed a number of GRB host
galaxies which were both observed at optical and NIR wavelengths, but
only detected in the visible.  This can be explained from the fact
that the GRB hosts display blue colours (see Sect.\,\ref{sec:colours})
and that, for the faintest sources at $R$\,$\sim$\,26--29, optical
deep observations are generally more sensitive than NIR images to
detect blue objects.  It is also the reason why the scatter in the
optical magnitudes appears larger than in the $K$--band.

In this ``$R$--z'' diagram, we have also indicated the $R$~magnitudes
of optically-selected galaxies from the Caltech Faint Galaxy Redshift
Survey \citep{Hogg00} and the Hubble Deep Field
\citep{Fernandez_Soto99}.  For the latter, the $R$--band photometry
was derived from the V and I magnitudes of the catalog assuming a
linear interpolation between the mean wavelengths of the V-band (F606
WFPC filter, 6031\AA) and the I-band (F801 WFPC filter, 8011\AA).  The
conversion from the standard AB magnitudes to the Vega system used
throughout this paper was obtained using the calibrations of
\citet{Fukugita95} and \citet{Allen00}.

Again, the GRB hosts in the visible appear just typical of the other
optically-selected galaxies at high redshift (see
Fig.\,\ref{fig:r_z}).  Yet, it is worth mentioning a particular
feature of the GRB host sample, which is clearly apparent in this
``$R$--z'' diagram. Whereas most of field sources at
$R$\,mag~$\gtapp$~25 have a redshift only determined with {\it
photometric\,} techniques, the GRB hosts have an accurate {\it
spectroscopic\,} redshift identification.  These redshifts were
derived using the emission and/or absorption features detected in the
X-ray/optical spectra of GRBs and their afterglows.  Such a method is
independent of the GRB host luminosities, and only depends on the
possibility to rapidly perform spectroscopy of the GRB transient
before it has begun to fade.  This advantage of GRBs for the selection
of high-z sources lies in stark opposition with the deep survey
approach.  Note that it is particularly apparent in the ``$R$--z''
diagram, but it is not that exceptional at NIR wavelengths (see
Fig.\,\ref{fig:k_z}). As mentioned previously, it is due to the blue
colours of GRB hosts, which thus allow the faintest of these hosts
detected in the $K$--band to be spectroscopically observed in the
visible.

\subsection{Colours}
\label{sec:colours}

So far, the various works related to the understanding of the high
redshift Universe have made an extensive use of the integrated $R-K$
colours of galaxies as an indicator of their nature.  These colours
provide indeed a crucial information on the importance of the old
stellar populations --- as traced by the NIR emission --- relative to
the contribution of young stars dominating the optical light. For
example, unobscured star-forming galaxies are typically blue objects
($R-K$\,$\sim$\,2--3) while old elliptical sources at z\,$\gtapp$\,1
exhibit extremely red colours ($R-K$\,$\gtapp$\,5). Furthermore, large
$R-K$ colours in distant sources can also suggest dust obscuration,
and may thus sign-post powerful dust-enshrouded starburst galaxies.

\begin{figure}[htbp]  
\resizebox{\hsize}{!}{\includegraphics{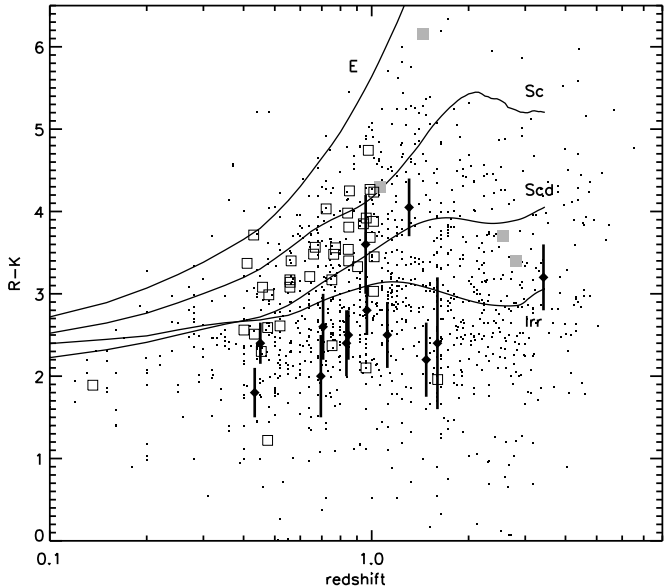}}
  \caption{Observed $R-K$ colours versus redshift for the sample of
  GRB host galaxies listed in Table\,\ref{tab:colours} ({\it filled
  diamonds}).  The estimated uncertainties are indicated with vertical
  solid lines.  The colours and redshifts for optically-selected field
  sources ({\it dots}) were derived from the HDF source catalog of
  \citet{Fernandez_Soto99}.  Solid curves indicate the observed
  colours of local E, Sc, Scd and Irr galaxies if they were moved back
  to increasing redshifts (see text for explanations).  We also
  indicated the colours of the ISO sources from the HDF ({\it open
  squares}) and those of SCUBA galaxies with confirmed redshifts ({\it
  filled squares}). See Sect.\,\ref{sec:colours} for references.  }
\label{fig:colours}
\end{figure}

In Table\,\ref{tab:colours}, we indicate the observed $R-K$ colours
for the $K$--selected sub-sample of GRB host galaxies.  The
corresponding diagram showing these colours versus redshift is
presented in Fig.\,\ref{fig:colours}.  As in the ``$K$--z'' relation
displayed in Fig.\,\ref{fig:k_z}, we also plotted the $R-K$ colours of
optically-selected sources taken from the HDF
\citep{Fernandez_Soto99}, and those of the ISO and SCUBA sources
already considered in the previous section. The $R$~magnitudes of the
ISO galaxies from the CFRS were derived using an interpolation between
the V and I magnitudes of \citet{Flores99}.  Those of the HDF ISO
detections and SCUBA sources were taken from the papers mentioned in
Sect.\,\ref{sec:mag_obs}.

In this diagram, we also indicate the hypothetical colours of typical
 present-day galaxies if they were moved to higher redshift assuming
 no evolution of their physical properties. These galactic templates
 were chosen to be mostly representative of the local Hubble sequence,
 and include both early-type (E/Sc) and late-type (Scd/Irr)
 sources. To compute the evolution of their $R-K$ colours with
 redshift, we used the optical/NIR templates of \citet{Mannucci01} for
 the E and Sc types, and the optical Scd and Irr SEDs of
 \citet{Coleman80}.  For the latter, the extrapolation to the
 near-infrared was derived using the NIR portion of the Mannucci et
 al. Sc template. This decision was justified from the prescriptions
 of \citet{Pozzetti96}, which show that the NIR continuum emission of
 {\it dust-free\,} galaxies longward of 1\,$\mu$m always appears
 dominated by the same stellar populations, and therefore does not
 vary much from one type to another.

As it can be seen in Fig.\,\ref{fig:colours}, the GRB hosts exhibit
rather blue colours that are typical of the faint blue galaxy
population in the field at z $\sim$ 1.  Besides, we note that most of
them appear even bluer than the colours predicted from the SED of
local irregular galaxies. This is similar to what has been already
noticed for a large fraction of blue sources detected in the optical
deep surveys \citep{Volonteri00}.  Such blue colours originate from
the redshifted blue continuum of the OB stars found in HII
regions. They are characteristic of unobscured star-forming galaxies,
which is not surprising in the scenario linking GRBs to massive star
formation. It is moreover in full agreement with the results of
\citet{Sokolov01} who found that the optical SEDs of the GRB host
galaxies are consistent with those of the blue starbursts observed in
the nearby Universe.

\subsection{Absolute K magnitudes}
\label{sec:mag_abs}

\begin{figure}[htbp]  
\resizebox{\hsize}{!}{\includegraphics{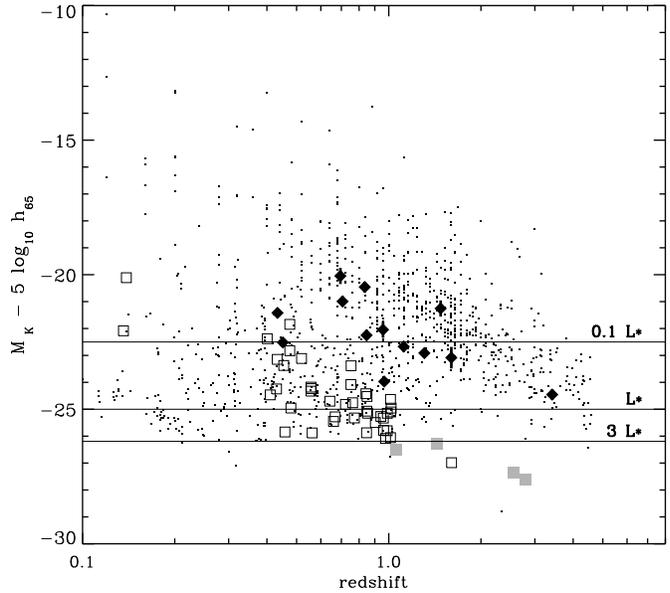}}
\caption{Absolute $K$ magnitudes of the GRB host galaxies listed in
Table\,\ref{tab:colours}, compared to optically/NIR-selected field
sources and ISO/SCUBA galaxies in a $\Lambda$CDM Universe
($\Omega_m$\,=\,0.3 and $\Omega_\lambda$\,=\,0.7). Legend, photometry
and redshift catalogs are similar to those used for
Fig.\,\ref{fig:k_z}, except for the NIR-selected objects with {\it
spectroscopic\,} redshifts which are also indicated by dots in this
plot. Horizontal lines indicate the magnitudes of 0.1\,L$_*$, L$_*$
and 3\,L$_*$ galaxies assuming M$_*$\,=\,--25.  }
\label{fig:k_abs_z}
\end{figure}  

We computed the absolute $K$ magnitudes for the sample listed in
Table\,\ref{tab:colours}, using the optical and NIR galaxy templates
described in Sect.\,\ref{sec:colours} to derive the k-corrections.
For all but two sources, the latter were obtained assuming an SED
typical of Irr-type objects as suggested by their blue $R-K$ colours
(see Sect.\,\ref{sec:colours}).  In the case of the GRB\,990506 and
GRB\,970828 hosts, we rather used an Scd-type template as indicated by
their redder SEDs (see Fig.\,\ref{fig:colours}).  Luminosity distances
were computed assuming a $\Lambda$CDM Universe with $\Omega_m$\,=\,0.3
and $\Omega_\lambda$\,=\,0.7.  We parametrized the Hubble constant
using h$_{65}$\,=\,H$_0$\,(km\,s$^{-1}$\,Mpc$^{-1}$)\,/\,65.

The absolute $K$ magnitudes are reported in Table\,\ref{tab:colours}
and illustrated in the Hubble diagram of Fig.\,\ref{fig:k_abs_z}.
Again, we also compared the GRB host galaxies with other field sources
quoted from the catalogs mentioned in Sect.\,\ref{sec:mag_obs}.  For
the galaxies of the HDF, the k-corrections used to compute these
magnitudes were derived assuming the best spectral type estimations of
\citet{Fernandez_Soto99}.  For the lower-redshift samples taken from
the NIR Hawaii Surveys and for the ISO sources, we arbitrarily assumed
an Irr galaxy template relying on the fact that k-corrections are
hardly type-dependent up to z\,$\sim$\,1.5 in $K$. The absolute $K$
magnitudes of the SCUBA galaxies were determined assuming the SED
suggested by their $R-K$ colour (see Fig.\,\ref{fig:colours}).

We also indicated the absolute magnitudes of galaxies with
luminosities of 0.1\,L$_*$ ($M\,=\,-22.5$), L$_*$ ($M\,=\,-25$) and
3\,L$_*$ ($M\,=\,-26.2$), assuming $M_*\,=\,-25$.  This value was
roughly estimated from the Schechter parametrizations of the $K$--band
luminosity function for high redshift galaxies, taken from
\citet{Cowie96}\footnote{$M_* = -25 + 5
\log_{10}$\,[H$_0$\,(km\,s$^{-1}$\,Mpc$^{-1}$)\,/\,50], $\alpha =
-1.3$} and \citet{Kashikawa03}\footnote{$M_* = -25.9 + 5
\log_{10}$\,[H$_0$\,(km\,s$^{-1}$\,Mpc$^{-1}$)\,/\,50], $\alpha =
-1.35$}.  Both were determined assuming a {\it matter-dominated\,}
Universe with $\Omega_m$\,=\,1.  Nevertheless, the differences in
comoving distance between a Universe with
H$_0$\,=\,50\,;\,$\Omega_m$\,=\,1\, and one characterized by
H$_0$\,=\,65\,;\,$\Omega_m$\,=\,0.3\,;\,$\Omega_\lambda$\,=\,0.7\,
imply absolute magnitude variations of only $\Delta$m\,=\,0.4 on the
0.7\,$\leq$\,z\,$\leq$\,3 redshift range. Therefore, it should not
significantly affect our qualitative comparison.

With a median--averaged $\overline{M_K}$\,=\,--22.25 (corresponding to
$\overline{L}\,\sim\,0.08\,L_*$), the GRB host galaxies are
significantly sub-luminous in the $K$--band.  We also note a large
difference with the luminosities of massive starbursts probed with ISO
and SCUBA, as it was already noticed in Sect.\,\ref{sec:mag_obs}.
Since the NIR emission of galaxies gives a good indication on their
mass, the low $K$--band luminosities of GRB hosts indicate that GRBs,
so far, were not observed toward massive objects.

\section{Discussion} 
\label{sec:discuss}

Our analysis described in Sect.\,\ref{sec:results}\, indicates that
the GRB host galaxies are characterized by rather blue colours (see
Fig.\,\ref{fig:colours}), sub-L$_*$ luminosities (see
Fig.\,\ref{fig:b_abs}) and low masses (see Fig.\,\ref{fig:k_abs_z}).
Their morphology is moreover consistent with that of compact,
irregular or merging systems \citep{Bloom02a}. Their spectra clearly
exhibit prominent emission lines such as [OII], [NeIII] and the Balmer
hydrogen lines \citep[e.g.,][]{Djorgovski98,LeFloch02a}, and their
optical SED is similar to that of starburst galaxies observed in the
local Universe \citep{Sokolov01}.  All together, our results and those
already published in the literature are therefore in agreement with
GRBs tracing star-forming sources at cosmological distances.

We may now wonder whether the GRB-selected objects are actually
representative of the {\it whole ensemble\,} of starburst galaxies at
high redshift, or in other words, whether GRBs can really be used as
unbiased probes of star formation.

\subsection{Is there any bias in the current sample of GRB hosts\,?}

Our results indicate that the GRB host galaxies significantly differ
from the luminous and dusty starbursts which were discovered in the
infrared and submillimeter deep surveys with ISO and SCUBA.  This is
clearly illustrated in Fig.\,\ref{fig:k_abs_z} which shows that,
contrary to the GRB hosts, these dusty star-forming objects are also
luminous at optical and NIR wavelengths. As can be seen in
Fig.\,\ref{fig:colours}, they moreover appear statistically much
redder, indicating either the presence of underlying old stellar
populations or an evidence for dust obscuration.  Is this distinction
between these galaxies and the GRB hosts simply due to the limited
size of our sample, or does it reveal other biases affecting the
GRB-selection\,?

The ISO starbursts have been shown to resolve $\sim$\,50\% of the
total energy produced by the Cosmic Infrared Background
\citep[CIRB,][]{Elbaz02b}. They trace therefore a significant fraction
of the global activity of star formation which occured in the
Universe.  Within the 0.5\,$\ltapp$\,z\,$\ltapp$\,1.5 redshift range
where the ISO sources and most of the GRB host galaxies are located,
we estimated the fraction of GRBs which should be observed toward
these dusty galaxies assuming GRBs trace the star formation.  To this
purpose, we assumed that the respective contributions to the CIRB and
the Optical Extragalactic Background (OEB) produced at this epoch
roughly originate from two distinct populations of star-forming
sources, namely the faint blue galaxies and the luminous dusty
starbursts \citep[e.g.,][]{Rigopoulou02}.  Given that the CIRB and the
OEB are more or less equivalent in terms of bolometric luminosity
\citep[e.g.,][]{Hauser01}, and taking into account the contribution of
the ISO sources to the CIRB, we found that approximately 25\,\% of the
GRB host galaxies should belong to the class of infrared dusty
starbursts such as those detected with ISO.  Within the sub-sample of
GRB hosts observed in the $K$--band and located at
0.5\,$\ltapp$\,z\,$\ltapp$\,1.5, $\sim$\,5--6 sources would thus be
expected to exhibit $K$~luminosities $\gtapp$\,0.5\,L$_*$, while none
of them actually satisfies this criterion.

Further evidence supporting the lack of GRB-detections in reddened dusty
starbursts is suggested by the very blue colours of GRB hosts. It has
been recently shown that a significant fraction of the Extremely Red
Objects (EROs, $R-K$\,$\gtapp$\,5) should be composed of dust-reddened
sources responsible for a star formation density greater than the
estimates from UV--selected galaxies at z\,$\sim$\,1 \citep{Smail02b}.
With a similar argument as aforementioned, we would expect to
find several EROs among the GRB host sample, while all of the GRB host
galaxies display $R-K$ colours bluer than $\sim$\,4.

This lack of luminous (L $\gtapp$ L$_*$) and red ($R-K \gtapp 4$)
galaxies among the GRB hosts could be explained by the existence of
the so-called ``dark'' bursts.  Lacking counterparts at optical/NIR
wavelengths in spite of a rapid and deep search of afterglows during
the few hours following their detection at high energy, a fraction of
these bursts could be hidden behind optically-thick columns of dust
and gas, and thus would be obscured in the visible. Indeed, the
spatial scale of dust-enshrouded regions of star formation in luminous
infrared galaxies can easily reach $\sim$~1\,kpc \citep{Soifer01}.
Even if the beamed emission of GRBs can destroy dust grains on
distances up to $\sim$~100\,pc from the burst location
\citep{Fruchter01b}, the resulting column density on the GRB line of
sight would still be high enough to prevent the production of a
detectable afterglow in the visible.  Such GRBs could thus only be
observed via the emission of their afterglows in the X-rays and,
possibly, through their synchrotron emission at radio wavelengths.
Since most of the currently known GRB hosts were selected using GRB
optical transients, it may therefore indicate that the sample is
likely biased toward galaxies harbouring unobscured star-forming
activity.

In this hypothesis, we would have the rough picture in which most of
GRBs with detectable optical transients mainly probe the dust-free
starbursts hosted in sub-luminous blue galaxies, while the bursts
occuring within the most dusty sources appear optically dark.
Naturally, intermediate cases should also exist, as illustrated by the
VLA detection of the GRB\,980703 host galaxy
\citep{Berger01b}. Assuming that the radio/far-infrared correlation
still holds for high redshift sources, this host pinpointed by an
optical transient of a GRB occuring at z\,=\,0.97 could be indeed a
dusty galaxy luminous in the infrared.  In fact, we note that its
$R-K$ colour $\sim$\,2.8 and its absolute $K$~magnitude $\sim$\,--24.0
would be consistent with this source being similar to the NIR
counterparts of the ISO dusty starbursts (see Figs.\,\ref{fig:colours}
and~\ref{fig:k_abs_z}).  Other intermediate examples of dusty galaxies
probed with optically bright GRB afterglows were also reported from
the faint detections of the GRB\,000418 and GRB\,010222 hosts at
submillimeter wavelengths \citep{Berger01a,Frail02}.

A possible method to reliably probe the dusty star formation with GRBs
could be the use of optically-dark bursts yet harbouring detectable
afterglows at radio wavelengths. However, the observations of four
sources pinpointed with such optically-dark and radio-bright GRBs by
\citet{Barnard03} have not revealed these galaxies to be especially
bright in the submillimeter. These particular bursts do not seem
therefore to preferentially select obscured sources. This indicates
that GRBs occuring within dust-enshrouded star-forming regions could
probably be dark at both optical/NIR and radio wavelengths, which
might be understood if GRB radio transients can not be easily
generated within the densest environments of dusty galaxies
\citep{Barnard03}.

It is therefore likely that the census of dust-enshrouded star
formation with GRBs will require a follow-up of the bursts
characterized by both optically- and radio-dark transients. The use of
their X-ray afterglows will thus be needed to correctly localize these
GRBs on the sky.  To this purpose, the forthcoming GRB-dedicated SWIFT
mission will enable to derive the positions of hundreds of GRBs with a
sub-arcsecond error box from the sole detections of their afterglows
in the X-rays.  This should ultimately provide a
statistically-significant sample of star-forming galaxies selected
from high energy transients of GRBs, thus less affected by dust
extinction than the current sample of GRB hosts.  The study of these
sources with the {\it Space InfraRed Telescope Facility\,} (SIRTF)
will moreover allow to characterize their dust content by directly
observing the thermal emission of these galaxies in the mid-infrared.
Since the SIRTF instruments will be able to detect the rest-frame
infrared emission of dusty starbursts up to z\,$\sim$\,2.5, the
parallel use of SWIFT and SIRTF will therefore open new perspectives
to use GRBs as probes of the dusty star formation at high redshift.

On the other hand, we note that this apparent selection effect toward
blue and sub-luminous sources may simply reflect an intrinsic property
of the GRB host galaxies themselves. For example, GRBs could be
preferentially produced within young systems experiencing their first
episode of massive star formation, thus explaining the low mass of
their underlying stellar populations and their apparent blue colors. A
larger statistics and a better understanding of the possible
observational bias associated with the GRB selection, as previously
mentioned, will be however required to further investigate this
hypothesis.

\subsection{Are GRB hosts representative of the faint blue galaxy population at high redshift\,?}
\label{sec:metal}

In the previous section, we have argued that the current sample of GRB
hosts could be biased toward unobscured star-forming galaxies, and
that such a bias could be related to the existence of the dark bursts.
Since the long-duration GRBs are believed to trace the star-forming
activity at high redshift, this sample of dust-free GRB-selected
sources should be therefore more or less representative of the
population of faint blue galaxies which were discovered in the optical
deep surveys.

These faint blue sources in the field are indeed believed to produce
the bulk of the OEB \citep{Madau00} and to be responsible for most of
the unobscured star formation in the early Universe.  Since the
$B$--band emission is a good tracer of star-forming activity in
dust-free sources, the absolute B magnitude histogram of the GRB host
sample, in this case, should closely follow the function of the
$B$--band luminosity {\it weighted by luminosity} for blue galaxies at
high redshift.  The latter may indicate indeed, for a given bin of
magnitude, the relative fraction of {\it total\,} star formation to
which galaxies in this range of luminosity contributed {\it as a
whole\,}. It should be therefore proportional to the fraction of GRB
occurence emerging from such galaxies.

\begin{figure}[htbp]  
\resizebox{\hsize}{!}{\includegraphics{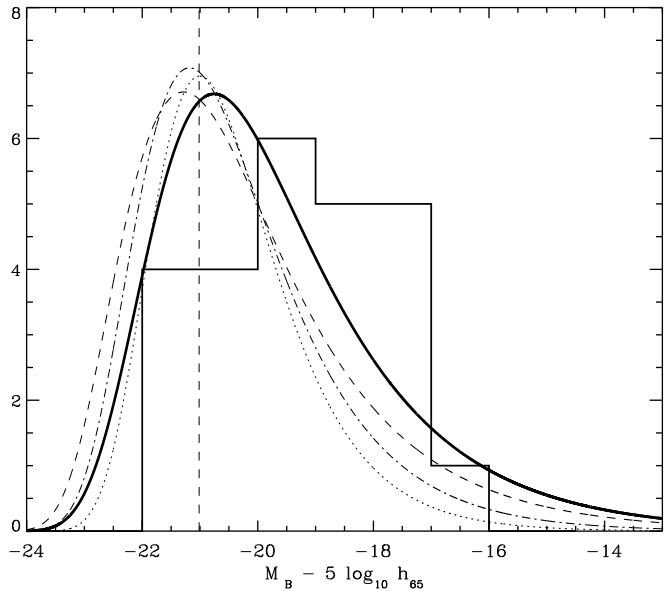}}
\caption{Histogram of the absolute B magnitudes for the sample of GRB
host galaxies, estimated following the method described in Appendix\,A
and assuming a $\Lambda$CDM Universe with $\Omega_m$\,=\,0.3 and
$\Omega_\lambda$\,=\,0.7.  Various functions of the $B$--band
luminosity {\it weighted by luminosity} at high redshift have been
overplotted in arbitrary units for comparison.  They were taken from
\citet{Wolf02} for the 0.8\,$\ltapp$\,z\,$\ltapp$\,1.2 redshift range
({\it dashed line\,}), and \citet{Kashikawa03} for sources at
1\,$\ltapp$\,z\,$\ltapp$\,1.5 ({\it dotted line\,}) and
1.5\,$\ltapp$\,z\,$\ltapp$\,2 ({\it dashed-dotted line\,}).  The thick
solid line is also taken from \citet{Wolf02} but restricted to
galaxies bluer than Sbc type objects.  The vertical dashed line
depicts the M$_*$ parameter of this last Schechter parametrization.}
\label{fig:b_abs}
\end{figure}

Such a comparison is shown in Fig.\,\ref{fig:b_abs}. We computed the
absolute magnitudes of the GRB host galaxies in the {\it rest-frame\,}
$B$--band following the method described in Appendix\,A.  To estimate
the contribution of distant sources to the overall star-forming
activity at similar redshifts, we used the results of the COMBO-17
Survey by \citet{Wolf02}, who derived the Schechter-parametrized
luminosity functions for various types of galaxies up to z\,$\sim$1.2,
in a $\Lambda$CDM Universe with $\Omega_m$=0.3 and
$\Omega_{\lambda}$=0.7.  We also used the observations of
\citet{Kashikawa03} from the Subaru Deep Survey who constrained the
global function of the $B$--band luminosity for sources up to
z\,$\sim$\,3.5, in a flat Universe with $\Omega_m$\,=\,1.  As
explained in Sect.\,\ref{sec:mag_abs}, the comoving distance
variations between the two different cosmologies do not affect that
much our comparisons.  Assuming $M_{B*} \sim -21 + 5 \log_{10} h_{65}$
for the blue galaxies at z\,$\sim$\,1 \citep{Wolf02}, it is clear
that the GRB host galaxies are sub-luminous sources in the
$B$--band.  There is however an apparent and surprising trend for the
GRB hosts to be, on average, even less luminous than the blue galaxies
which mostly contributed to the energy density in the rest-frame
$B$--band at z\,$\sim$\,1. This apparent shift is unlikely due to the
weak constrain which has been obtained so far on the faint end slope
(usually refered as the parameter $\alpha$ in the Schechter
parametrization) of the $B$--band luminosity function at high
redshift.  There are indeed noticeable discrepancies in this slope
between the results of Kashikawa et al. $(-1.2\,\ltapp \alpha
\ltapp\,-0.9)$ and those by Wolf et al. $(-1.5\,\ltapp \alpha
\ltapp\,-1.3)$, but the implied variations on the $B$--band luminosity
function {\it weighted by luminosity} are not that significant (see
Fig.\,\ref{fig:b_abs}).  To quantify the observed shift, we performed
a Kolmogorov-Smirnov test on the data sets of GRB hosts and high
redshift sources bluer than Sbc type objects.  We obtained only a
rather small probability ($\sim$\,17\%) that the two distributions
originate from the same population of galaxies.  Could this shift be
due to intrinsic properties of GRB host galaxies, and reveal that only
particular environments favour the formation of GRB events\,?

In the ``collapsar'' scenario, GRBs are produced by the accretion of a
helium core onto a black hole resulting from the collapse of a
rapidly-rotating iron core.  Since a low metallicity in the stellar
enveloppe reduces the mass loss and inhibits the loss of angular
momentum by the star, the formation of GRBs could be favoured in
metal-poor environments \citep{MacFadyen99}.  As such, we could expect
GRBs to be preferentially observed toward starbursts with very low
luminosities.  Interestingly, evidence for a low-metallicity host
galaxy has been in fact recently reported toward the X-Ray Flash
XRF\,020903 \citep{Chornock02}.  The influence of such intrinsic
parameters could therefore not only explain the trend observed in
Fig.\,\ref{fig:b_abs}, but it would also provide further arguments for
the lack of GRB host detections toward luminous reddened starbursts as
discussed in the previous section.

Of course, one must remain very cautious regarding this
interpretation, because of the small size of our sample.  Moreover,
the influence of metallicity in the formation of a GRB should be only
localized in the close vicinity of the burst, whereas important
gradients in the chemical composition of galaxies are commonly
observed.  However, we note that such gradients of metallicity are not
present in local dwarfs \citep[see][ and references
therein]{Hidalgo01}, which suggests that the average metallicity
observed toward the sub-luminous GRB hosts should give a good
estimation of the chemical properties characterizing the environments
where the GRBs occured.  A detailed investigation of the gas
metallicity within GRB host galaxies compared to other
optically-selected sources has never been performed so far.  Such a
study will be definitely required to better address this issue.

\section{Conclusions}
Using our $K$--band observations of GRB host galaxies in combination
with other optical and NIR data of the literature, we conclude that:

1) Most of the GRB hosts discovered so far belong to the population of
faint and blue star-forming galaxies at high redshift. They have low
masses as suggested by their faint luminosity in the near-infrared.
They are also sub-luminous sources at optical wavelengths. Most of
them are characterized by intrinsic $R-K$ colours even bluer than
those displayed by the starburst galaxies observed in the nearby
Universe.

2) The lack of GRB detection toward luminous starbursts and/or
reddened sources such as those observed in the infrared and
submillimeter deep surveys seems to indicate a possible bias of the
currently-known GRB host sample against this type of objects.  This
could be explained by the fact that the selection of GRB host
galaxies, so far, had to rely on the identification of optical GRB
afterglows likely probing unobscured star-forming galaxies. The
follow-up of optically-dark and radio-dark GRBs, and the use of their
X--ray afterglows to obtain their localization with a sub-arcsecond
error box will be likely necessary to probe dust-enshrouded
star-forming galaxies in the early Universe with these particular
phenomena.  On the other hand, the hypothesis that such a bias of
selection is purely intrinsic to the GRB host properties can not be
rejected, assuming that GRBs preferentially occur within young and
blue starbursts.

3) The observed GRB host galaxies seem to be statistically less
luminous than the faint blue sources which mostly contributed
to the $B$--band light emitted at high redshift. This could reveal an
intrinsic bias of the GRB selection toward star-forming regions with
very low luminosities, and might be explained taking account of
particular environmental properties (e.g., metallicity) favouring the
formation of Gamma-Ray Burst events. In this context, this could also
indicate that GRBs can not be used as unbiased probes of star
formation.  A larger statistics of the GRB host absolute $B$
magnitudes and a detailed study of the chemical composition of the gas
within GRB host galaxies will be however required to further confirm
this result.

\label{sec:conclusion}

\begin{acknowledgements}  
We would like to specially thank the teams of the NTT and ESO--3.6m at
La Silla for their kind and efficient assistance during the
observations in visitor mode.  We have also appreciated the work of
the ESO-Paranal and Gemini-North staff regarding the acquisition of
the VLT and Gemini data in service mode.  We acknowledge F.\,Mannucci
for publicly providing his optical/NIR spectral templates via a
user-friendly web interface, as well as A.\,Fern\'andez-Soto for
maintaining his HTML access to the HDF photometric redshift catalog.
We are grateful to C.\,Lidman for his advice in the NIR data reduction
techniques, as well as H.\,Aussel, L.\,Cowie, D.\,Elbaz, R.\,Chary,
F.\,Combes, E.\,Feigelson and C.\,de~Breuck for useful discussions
related to this work.  We finally thank our referee, S.A. Eales, for
interesting comments and suggestions on this paper.  This research
project was partially supported by CONICET/Argentina and Fundacion
Antorchas. DM is supported by FONDAP Center for Astrophysics 15010003.
\end{acknowledgements}

\appendix

\section{Absolute B magnitudes of GRB host galaxies}

The dereddened $R$ magnitudes of the GRB host galaxies with determined
redshifts are given in Table\,\ref{tab:r_mag}. They were estimated in
the Vega system taking account of {\it most of\,} the published papers
and GRB Coordinates Network circulars directly or indirectly reporting
on optical observations of fading GRB afterglows and their hosts.
When several $R$ magnitudes of a given source were available in the
literature, the various measurements were weighted according to their
photometric uncertainty, and subsequently averaged to get a final
homogenized value. In some cases, we also relied on the host
contribution derived from the fit of the $R$--band optical transient
light-curve when the latter was clearly well constrained.  The $R$
magnitude of the GRB\,990506 host galaxy was measured from an
$R$--band image that we obtained using the EFOSC2 instrument on the
ESO 3.6-m telescope at La Silla.

The redshifts given in Table\,\ref{tab:r_mag} have also been taken
from the literature. In most cases, they were determined from emission
lines directly observed in the spectra of the hosts. For the other
sources, they were derived as the redshifts of the furthest absorbing
medium observed in absorption within the spectra of the GRB optical
transients.  We made the assumption that the first interstellar medium
illuminated by the background afterglow is indeed that of its host
galaxy itself. We note that this hypothesis has been confirmed in
several cases where the derived redshift could have been confirmed
with emission lines from the host.

These redshifts and $R$ magnitudes were subsequently used to derive
 the absolute $B$ magnitudes given in Column~(1) of
 Table\,\ref{tab:r_mag}, assuming a $\Lambda$CDM Universe with
 $\Omega_m = 0.3$, $\Omega_{\lambda} = 0.7$.  For each host, the
 k-correction for the $R$--filter and rest-frame $B-R$ colour used for
 this computation were estimated taking account of the type of SED
 suggested by its $R-K$ and/or optical colours when available (see
 Table\,\ref{tab:colours} and Fig.\,\ref{fig:colours}), otherwise
 assuming a blue continuum with a spectral slope $F_{\nu} \varpropto
 \nu^{-1}$.

To better establish the validity of our results, we also estimated,
for most of the hosts, the absolute $B$ magnitudes from the observed
flux density at the redshifted $B$--band wavelength. For each case,
this flux density was derived interpolating the various broad-band
filter magnitudes given in the literature (see Table\,\ref{tab:r_mag}
for references) including the $K$~magnitudes given in
Table\,\ref{tab:colours}.  The final results are indicated in
Column~(2) of Table\,\ref{tab:r_mag}. To compare the two methods, we
computed the difference between the estimations given in the two
columns, and found a mean value {\tt <\,}M$_{\rm B(1)}$\,--\,M$_{\rm
B(2)}$\,{\tt >}\,=\,0.07 and a dispersion $\sigma$\,(M$_{\rm
B(1)}$\,--\,M$_{\rm B(2)}$)\,=\,0.18.

\begin{table*}[htbp]

\caption{Optical ($R$ and $B$--band) properties of GRB host galaxies}
\begin{center}
\begin{tabular}{ccccccllccc}
\hline \hline
      &      &  \multicolumn{2}{c}{Redshift}  & & \multicolumn{3}{c}{Photometry} & & 
\multicolumn{2}{c}{ $M_B + 5 \log_{10} h_{65}~(\P)$} \\ 
\cline{3-4} \cline{6-8} \cline{10-11 }\\
Source & GRB & z & Ref. & & \multicolumn{1}{c}{$E(B-V)^{\dag}$} & 
\multicolumn{1}{c}{$R$ mag} &  \multicolumn{1}{l}{References} & & (1) & (2)\\
\hline \\
GRB J194641.9--193605  & 020813 & 1.25 & 1  & & 0.109 & 24.70 $\pm$ 0.20   	  & 2              & & -19.30 &        \\
GRB J151455.8--192454  & 020531 & 1.00 & 3  & & 0.140 & 22.05 $\pm$ 0.20   	  & 4              & & -21.35 &        \\
GRB J135803.1--312222  & 020405 & 0.69 & 5  & & 0.050 & 20.90 $\pm$ 0.20   	  & 5              & & -21.50 &        \\
GRB J111518.0--215656  & 011211 & 2.14 & 6  & & 0.036 & 24.90 $\pm$ 0.30   	  & 7              & & -20.55 &        \\
GRB J113429.6--760141  & 011121 & 0.36 & 8  & & 0.508 & 24.60 $\pm$ 0.40   	  & 9              & & -16.15 & -16.25 \\
GRB  J225559.9+405553  & 010921 & 0.45 & 10 & & 0.145 & 21.45 $\pm$ 0.15   	  & 10, 11         & & -19.75 & -19.95 \\
GRB  J145212.5+430106  & 010222 & 1.48 & 12 & & 0.023 & 25.70 $\pm$ 0.15   	  & 13, 14         & & -18.75 & -18.50 \\
GRB  J170409.7+514711  & 000926 & 2.04 & 15 & & 0.024 & 24.80 $\pm$ 0.10   	  & 15             & & -20.50 &        \\
GRB  J021834.5+074429  & 000911 & 1.06 & 16 & & 0.120 & 25.10 $\pm$ 0.10   	  & 16             & & -18.80 & -18.85 \\
GRB  J122519.3+200611  & 000418 & 1.12 & 17 & & 0.033 & 23.80 $\pm$ 0.20   	  & 18, 19         & & -19.90 & -19.85 \\
GRB  J162018.6+292636  & 000301 & 2.04 & 20 & & 0.052 & 27.85 $\pm$ 0.30   	  & 21             & & -17.45 &        \\
GRB J015915.5--403933  & 000210 & 0.85 & 22 & & 0.017 & 23.45 $\pm$ 0.10   	  & 22             & & -19.50 & -19.50 \\
GRB  J050931.3+111707  & 991216 & 1.02 & 23 & & 0.633 & 25.30 $\pm$ 0.20   	  & 24             & & -18.15 &        \\
GRB  J163353.5+462721  & 991208 & 0.71 & 25 & & 0.016 & 24.30 $\pm$ 0.20   	  & 25, 26         & & -18.30 & -18.30 \\
GRB J223153.1--732429  & 990712 & 0.43 & 27 & & 0.032 & 21.85 $\pm$ 0.15   	  & 28, 29, 30     & & -19.35 & -19.50 \\
GRB J050954.5--720753  & 990705 & 0.84 & 31 & & 0.122 & 21.80 $\pm$ 0.10   	  & 31, 32, 33     & & -21.65 & -21.75 \\
GRB J133807.1--802948  & 990510 & 1.62 & 27 & & 0.118 & 27.50 $\pm$ 0.30$^\ddag$  & 34             & & -17.20 &        \\
GRB J115450.1--264035  & 990506 & 1.31 & 17 & & 0.065 & 25.50 $\pm$ 0.20          & this work      & & -19.75 & -19.45 \\
GRB  J152530.3+444559  & 990123 & 1.60 & 35 & & 0.016 & 24.30 $\pm$ 0.40          & 26, 36, 37, 38 & & -20.40 & -20.05 \\
GRB  J235906.7+083507  & 980703 & 0.97 & 39 & & 0.058 & 22.40 $\pm$ 0.20          & 26, 40, 41, 42 & & -20.90 & -20.80 \\
GRB  J101757.8+712725  & 980613 & 1.10 & 43 & & 0.090 & 25.80 $\pm$ 0.10          & 44             & & -17.85 & -17.90 \\
GRB  J115626.4+651200  & 971214 & 3.42 & 45 & & 0.016 & 25.60 $\pm$ 0.20          & 26, 45, 46     & & -21.30 & -21.30 \\
GRB  J180831.6+591851  & 970828 & 0.96 & 47 & & 0.038 & 25.10 $\pm$ 0.30          & 47             & & -18.85 & -18.35 \\
GRB  J065349.4+791619  & 970508 & 0.83 & 48 & & 0.049 & 25.10 $\pm$ 0.20          & 26, 48, 49     & & -17.85 & -17.75 \\
GRB  J050146.7+114654  & 970228 & 0.69 & 50 & & 0.234 & 24.60 $\pm$ 0.20          & 50, 51, 52     & & -17.85 & -17.80 \\

\hline \hline
\end{tabular}\\
\label{tab:r_mag}
\end{center}
{\bf Notes:}

\hspace{.2cm} $\dag$ : foreground Galactic extinction. For all cases
excepted GRB\,990705, it has been estimated from the DIRBE/IRAS dust

\hspace{.64cm} maps of \citet{Schlegel98}. For GRB\,990705
which occured behind the Large Magellanic Cloud, we used
the results

\hspace{.64cm}  of \citet{Dutra01}.

\hspace{.2cm}
$\ddagger$ : derived from the $V$ magnitude of \citet{Fruchter00c} assuming a spectral
slope $F_{\nu} \varpropto \nu^{-1}$.

\hspace{.15cm}
$\P$ : the absolute $B$ magnitudes were derived 
assuming a $\Lambda$CDM Universe with $\Omega_m = 0.3$ and $\Omega_{\lambda} = 0.7$.
Column~(1) gives the

\hspace{.64cm} estimations which were obtained by applying proper k-corrections
and $B-R$ colours to the observed $R$ magnitudes. The 

\hspace{.64cm} results given in Column~(2) were derived
from the observed flux density at the redshifted $B$--band wavelength.
\hspace{.02cm}

\vspace{0.1cm}

{\bf References:} 
(1) \citealt{Price02a}~; 
(2) \citealt{Levan02}~; 
(3) \citealt{Kulkarni02}~;
(4) \citealt{Fox02}~; 
(5) \citealt{Price02b}~;
(6) \citealt{Holland02}~;
(7) \citealt{Burud01}~; 
(8) \citealt{Garnavich02}~;
(9) \citealt{Bloom02b}~;
(10) \citealt{Price02}~:
(11) \citealt{Park02}~; 
(12) \citealt{Jha01}~; 
(13) \citealt{Frail02}~;
(14) \citealt{Fruchter01c}~;
(15) \citealt{Castro02}~; 
(16) \citealt{Price02c}~;
(17) \citealt{Bloom02c}~;
(18) \citealt{Metzger00}~; 
(19) \citealt{Klose00}~;
(20) \citealt{Jensen01}~;
(21) \citealt{Fruchter01}~; 
(22) \citealt{Piro02a}~;
(23) \citealt{Vreeswijk99a}~; 
(24) \citealt{Vreeswijk00}~; 
(25) \citealt{Castro-Tirado01}~:
(26) \citealt{Sokolov01}~; 
(27) \citealt{Vreeswijk01}~;
(28) \citealt{Fruchter00a}~; 
(29) \citealt{Hjorth00}~;
(30) \citealt{Sahu00}~;
(31) \citealt{LeFloch02a}~; 
(32) \citealt{Saracco01b}~;
(33) \citealt{Holland00c}~; 
(34) \citealt{Fruchter00c}~; 
(35) \citealt{Kulkarni99}~;
(36) \citealt{Fruchter99a}~; 
(37) \citealt{Holland99}~; 
(38) \citealt{Bloom99}~; 
(39) \citealt{Djorgovski98}~;
(40) \citealt{Bloom98b}~;
(41) \citealt{Holland01}~; 
(42) \citealt{Vreeswijk99}~; 
(43) \citealt{Djorgovski00}~; 
(44) \citealt{Hjorth02}~; 
(45) \citealt{Kulkarni98}~;
(46) \citealt{Odewahn98}~;
(47) \citealt{Djorgovski01a}~; 
(48) \citealt{Bloom98a}~; 
(49) \citealt{Fruchter00b}~; 
(50) \citealt{Bloom01b}~;
(51) \citealt{Fruchter99b};
(52) \citealt{Galama00}.
\end{table*}

\end{document}